\documentclass[aps,showpacs,twocolumn,amsmath,amssymb,floatfix]{revtex4}
\usepackage{amsthm}
\usepackage{latexsym}
\usepackage{amsfonts}
\usepackage{bbm,dsfont}
\usepackage{graphicx,color}
\newcommand{\R}{\mathbb{R}}

\newcommand{\Id}{\mathbb{I}}

\newcommand{\tr}[1]{\mathrm{Tr}\left[ {#1} \right]} 
 
\newcommand{\ket}[1]{\left\vert {#1} \right\rangle}

\newcommand{\ketbra}[2]{\left\vert {#1} \right\rangle\left\langle {#2} \right\vert}

\newcommand{\ft}{\color{black}}
\newcommand{\mgap}{\color{black}}
\begin{document}
\title{Exact and approximate solutions for the quantum minimum-Kullback-entropy 
estimation problem}
\author{Carlo Sparaciari}\email{carlo.sparaciari@studenti.unimi.it}
\affiliation{Dipartimento di Fisica dell'Universit\`a degli Studi di Milano,
I-20133 Milano, Italia.}
\author{Stefano Olivares}\email{stefano.olivares@mi.infn.it}
\affiliation{Dipartimento di Fisica dell'Universit\`a degli Studi di Milano,
I-20133 Milano, Italia }
\affiliation{CNISM UdR Milano Statale, I-20133 Milano, Italy}
\author{Francesco Ticozzi}\email{ticozzi@dei.unipd.it}
\affiliation{Dipartimento di Ingegneria dell'Informazione, 
Universit\`a di Padova, I-35131 Padova, Italia}
\affiliation{Department of Physics and Astronomy, Dartmouth College, 6127 Wilder, Hanover, NH 03755 (USA).}
\author{Matteo G. A. Paris}\email{matteo.paris@fisica.unimi.it}
\affiliation{Dipartimento di Fisica dell'Universit\`a degli Studi di Milano,
I-20133 Milano, Italia}
\affiliation{CNISM UdR Milano Statale, I-20133 Milano, Italy}
\begin{abstract}
The minimum Kullback entropy principle (mKE) is a useful tool to
estimate quantum states and operations from incomplete data and 
prior information. In general, the solution of a mKE problem is
analytically challenging and an approximate solution has been proposed
and employed in different context. Recently, {\ft the form and a way to
compute the} exact solution for finite dimensional systems has been
found, and a question naturally arises on whether the approximate
solution could be an effective substitute for the exact solution, and in
which regimes this substitution can be performed. Here, we provide a
systematic comparison between the exact and the approximate mKE
solutions for a qubit system when average data from a single
observable are available. We address both mKE estimation of states and
weak Hamiltonians, and compare the two solutions in terms of state
fidelity and operator distance.  We {\ft find} that the approximate
solution is generally close to the exact one unless the initial state is
near an eigenstate of the measured observable.  
{\mgap Our results provide a rigorous justification for the use of the approximate
solution whenever the above condition does not occur, and extend its
range of application beyond those situations satisfying the assumptions used 
for its derivation.}
\end{abstract}
\date{\today}
\maketitle
\section{Introduction}
Let us consider a situation where a quantum system is prepared in a
known state and {\ft, after some time and unknown evolution, some 
measurements are performed in order to gather information on the final
state}. 
The {\ft exact} solution to this state estimation problem is provided by
quantum tomography, which however requires the measurement of a complete
set (i.e. a {\em quorum}) of observables \cite{Dar01}.  \par Measuring a
quorum of observables may be experimentally challenging, or {\ft require
too many resources}, and therefore it is worth {\ft exploring} the case
where the set of observables that can be measured on the system is
incomplete \cite{LNP}.  In this case, we cannot obtain complete
information about the state of the system from the outcomes of these
measurements, i.e. the measurements are not fully determining the state
of the system.  We thus need some additional ingredient to fill the
information gap and single out a quantum state that is compatible with
the data, and with the information that is possibly available prior to
the measurements \cite{zhr97,zhr00}. 
\par 
When we have no {\em a priori} information, e.g. because the initial
state of the system is unknown or the interaction with the environment
is strong enough to wash out any initial information, the problem may be
attacked using the maximum entropy principle (ME) \cite{Jaynes,rmp13}. 
With
the ME we take all the states (density  matrices) compatible with the
evidences, i.e. reproducing the correct probabilities of the observed
data, and pick up the one maximizing the Von Neumann entropy. In this
way, the only knowledge about the state is that coming from the
measurements made on the system, without the addition of any unwanted
piece of extra information which is not available from the experimental
evidences \cite{Buz97,Zim08}.
\par 
On the other hand, there are several situation of interest where some
{\em a priori} information is indeed available, {\ft in the form of a
{\em a priori} state.} This may be due to some constraints imposed to
the physical preparation of the system, or to the fact that the coupling
of the system with the rest of the universe is weak, {\ft so that} the
state remains close to the initial preparation.  In these cases, the
minimum Kullback entropy (mKE) principle
\cite{Kull51,Shore80} 
provides an effective tool to {\ft
include this new ingredient in the solution} and to {\ft
complement the experimental data, thus allowing to obtain a unique
estimated state.} 
\par
The mKE principle has received attention in the recent years and has
been applied to both finite and infinite-dimensional systems
\cite{Geo06,Oli07,Tic10,Zor11}.  In particular, applications to
qubit and harmonic oscillator systems have been initially put forward
upon exploiting an approximate solution of the minimization problem
\cite{Oli07,Oli12}.  More recently, the analysis of the feasibility,
{\ft the form of the general solution and a method to compute it} has
been {\ft derived for finite dimensional systems} \cite{Ticozzi} and a
question naturally arises on {\ft how the approximate solution compares
to the exact one,} and in which regimes it could be convenient to
employ the former.  {\mgap
Indeed, our analysis is motivated by two relevant properties of 
the approximate solution: on the one hand it is given in a closed 
form which is more convenient for applications 
and, on the other hand, it may be applied to a larger class of a priori 
states, including those described by a density operator not having 
full rank.}
\par
This paper {\ft focuses on} qubit systems in situations where only {\ft the
average of a} single observable {\mgap can be accessed}.  
We consider the use of mKE
for estimation of states and for the characterization of weak
Hamiltonians, and compare the two solutions in terms of state fidelity
and operator distance respectively. We {\ft find} that the approximate
solution is generally close to the exact one unless the initial state is
near an eigenstate of the measured observable. {\mgap Our results
thus provide a rigorous justification for the use of the approximate
solution whenever the above condition does not occur.}
\par
The paper is structured as follows. In Sec.\ \ref{s:mke} we review the 
mKE principle for a qubit system where a single observable is measured,
and present both the approximate and the exact solutions to the 
mKE estimation problem. In Sec.\ \ref{s:cmp} a systematic comparison between the
approximate and the exact solution is performed in terms of fidelity, and 
the role of initial purity is discussed.
In Sec.\ \ref{s:ham} we address estimation of weak Hamiltonians by mKE 
and compare the approximate and the exact solutions in terms of operator
trace distance. Sec.\ \ref{s:out} closes the paper with 
some concluding remarks.
\section{The minimum Kullback entropy principle} \label{s:mke}
The quantum Kullback ({\ft Umegaki's}) relative entropy between two quantum states is
defined as \cite{Umegaki,Pet91,Hay01,Ved02}:
\begin{equation}
K( \rho | \tau) = \tr{\rho ( \log\; \rho  - \log\; \tau )}\,.
\end{equation}
As for its classical counterpart, the Kullback-Leiber divergence, 
it can be demonstrated that $0\leq K(\rho | \tau) < \infty$
when it is definite, i.e. when the support of the first state in the Hilbert
space is
contained in that of the second one.
In particular, $K(\rho | \tau )=0$ iff $\rho\equiv\tau$.
This quantity, though not defining  a proper metric in the Hilbert
space (it is not simmetric in its arguments), 
has been widely used in different fields of quantum information
\cite{Ved02,Sch02,Gal00,Qin08,ng1,ng2} because of its additivity properties 
and statistical meaning in state discrimination.
\par
Let us now consider a quantum system initially prepared in the state 
$\tau$ that, after some kind of evolution, unitary or not, is now in the final 
state $\rho$. In this case we have some prior information that we can 
regard as a bias towards $\tau$. Furthermore, when some observables are
measured, the  information achieved (e.g. their mean values or their full
probability distributions) gives some constraints about the state. 
The mKE principle, states that the best estimate for the 
state $\rho$ is then the density matrix that satisfies the constraints and,
at the same time, is somehow closer to the initial state $\tau$, i.e. 
minimize the quantum Kullback entropy, given the constraints.
The mKE principle allows one to take into account the available prior 
information as well as the new evidence coming from the data, 
while not introducing any other kind of spurious or unwanted piece of
information.
\par
The minimization can be done by Lagrange multipliers. If the 
constraint is given by the mean value of the observable $A$ (and
by the normalization), then the quantity that should be 
minimized is:
\begin{align} \label{Lagrange}
F( \rho , \lambda_1, \lambda_2 ) = 
K( \rho | \tau) \; & + \lambda_1 (\tr{\rho} - 1) 
\nonumber \\
& + \lambda_2 ( \tr{\rho A} - \langle A \rangle ),
\end{align}
$\lambda_k$ being the Lagrange multipliers.
{\ft Two approaches have been developed} to solve this mKE problem. The first 
is approximate and leads to analytic solutions in several cases,
e.g.\ when the final state is close to the initial one  
\cite{Bra96,Oli07}.
More recently, the general feasibility of this estimation problem and an
exact method has  been developed, valid when the quantum system
under investigation is finite dimensional \cite{Ticozzi}. 
{\mgap Having at disposal an exact solution allows us to assess the
approximate one and to individuate the situations where it may
safely apply instead of the exact one}.
In the following we
are going to systematically compare the two solutions for a qubit system
subjected to the measurement of a single observable.
\subsection{General Solution}
It is possible to show that, for finite-dimensional
Hilbert spaces, the minimum of the Eq.\ (\ref{Lagrange})
exists, is unique and {\ft continuous with respect to the data \cite{Ticozzi}}. {\ft After a suitable reduction of the problem to a subspace that ensures that the solution is full rank, and assuming that $\tau$ is full-rank on the same subspace,} the optimal solution of the
problem, when the only outcome of the measurement is the
mean value of observable $A$, is the following one:
\begin{equation} \label{exact}
\rho (\lambda_1 , \lambda_2 ) = e^{ \log{\tau} - \Id - \lambda_1 X_1 - \lambda_2 X_2}
\end{equation}
where $\lambda_1$, $\lambda_2$ are Lagrange multipliers, and
$X_1$, $X_2$ are the operators obtained from $\Id$ and $A$
through the Gram-Schmidt orthogonalization process.
{\ft Notice that if $\tau$ is not full rank the above formula does not return a valid density operator.}
In order to evaluate the Lagrange multipliers, the constraints
of normalization $\hbox{Tr}[\rho(\lambda_1, \lambda_2)]=1$ and 
mean value of $A$,
$\hbox{Tr}[\rho(\lambda_1, \lambda_2)\,A]=\langle A\rangle$ 
should be imposed.
\subsection{Approximate solution}
{\mgap Assuming that the evolution is not leading the system too
far away from its initial preparation, we can find an
approximate solution to the mKE estimation problem upon
writing the infinitesimal increment of the density
operator. More explicitly: in the Hilbert space of statistical 
operators, one considers an infinitesimal increment of the operator 
$\rho$ correspoding to the increment
of an arbitrary parameter $\lambda$. Upon assuming that increments
of the density operator are evaluated according to the Fisher metric
it is possible to introduce the 
differential equation \cite{Bra96}:
\begin{equation} \label{diff}
\frac{d\rho}{d\lambda} = -\frac{1}{2} \{ \rho , A - \langle A \rangle \}
\end{equation}
where $\{\ ,\ \}$ is the anticommutator. 
As already mentioned, the same equation can be obtained from Eq.\ (\ref{Lagrange}), when
the final state $\rho$ is close to the initial state $\tau$,
according to the Fisher metric. In turn,
the state $\rho$ obtained by integration of Eq.\ (\ref{diff})
is the approximate solution of the mKE problem with $\lambda$ playing
the role of a Lagrange multiplier.
}
\par This work focuses on statistical operator in the qubit space,
when the initial state is given and the only information
obtained from measurement is the mean value of the observable
$A$. The solution of the previous equation for this case
(notice that it is also correct for spaces with dimension larger than two) is:
\begin{equation} \label{approx}
\rho (\lambda ) = \frac{ e^{-A \lambda / 2} \tau e^{ -A \lambda / 2} }{ \tr{ \tau e^{ -A \lambda } } }
\end{equation}
where $\lambda$ can be found using the constraint $\tr{A \rho} = \langle
A \rangle $. 
{\mgap As mentioned above, the approximate solution $\rho(\lambda)$ may 
be computed also if $\tau$ is not full rank. In addition, we notice that 
$\rho(\lambda)$ has the same rank of the a priori state $\tau$.}
\par
{\mgap Since the approximate solution has been derived
assuming that evolved state is close to the initial one
\cite{Bra96}, one may expect that $\rho(\lambda)$ obtained from Eq.
(\ref{approx}) is not too far away from the a priori state $\tau$.
As we will show in the following, this is basically true in the case of
nearly pure initial states. Otherwise, when the initial state is
appreciably mixed, the approximate solution can, in fact, be far 
away from the initial preparation. On the other hand, also in these cases the
approximate solution is close to the exact one.
In other words, having at disposal the exact solution allows us to
assess the approximate one also outisde the assumptions made to derive
it, and to extend the regimes where it may be safely employed.
}
\section{Comparison between the exact and the approximate mKE estimates}
\label{s:cmp}
In this section the two solutions are compared, through the use of 
the fidelity, in order to establish whether, and in which conditions, 
the approximate solution can be considered as a good replacement for 
the exact one. The comparison is made for qubits systems.
\par
In particular, we address situations where a single observable {\ft $A$} is 
measured. The most general qubit observable may be written as
{\ft $$A = a_0 \Id + {\boldsymbol a} \cdot \boldsymbol \sigma\,,$$} 
where $a_0$ and $\boldsymbol a=(a_1,a_2,a_3)$ are real parameters
and $\boldsymbol \sigma=(\sigma_1, \sigma_2, \sigma_3)$ denote the
Pauli matrices vector. Without loss of generality it is always possible to 
perform a rotation and a scaling in order to rewrite the observable as
{\ft \begin{equation}\label{operatore}
U^\dag\, A\, U = A =\alpha \Id + \sigma_3
\end{equation}}
i.e. as a function of a single real parameter $\alpha$.
\par 
Once the rotation is made, we rewrite the general state of the 
qubit in the new reference as 
$$\tau = \frac{1}{1+\epsilon} (\ketbra{\psi}{\psi} 
+ \epsilon \ketbra{\psi^\perp}{\psi^\perp})\,,$$
where $$\ket{\psi} = \cos{\frac{\theta}{2}} \ket{0} 
+ e^{i \phi} \sin{\frac{\theta}{2}} \ket{1}$$
is the generic pure state and $\ket{\psi^\perp}$ its orthogonal
complement.  The parameter $\epsilon$ depends on the purity $\mu[\tau]$ of the 
initial state $\tau$, we have:
$$ \mu[\tau]=\frac{1+\epsilon^2}{(1+\epsilon)^2}\,,$$
with $\mu \in [1/2,1]$.
\par 
The parameters involved in this problem are five. The three parameters 
$\theta \in [0,\pi]$, $\phi \in [0,2 \pi)$ and $\mu \in [1/2,1]$ are needed 
to fully characterize the initial state, whereas $\alpha \in \R$ 
and $\langle \sigma_3 \rangle \in [-1,1]$ specify the measured
observable and its mean values respectively: $\langle A \rangle =
\alpha + \langle \sigma_3 \rangle$. Since we are dealing with qubit
measurements (which have two possible outcomes) the knowledge of the
mean value is equivalent to that of the full distribution.
\par 
Once {\ft we fix} both $\tau$ and $A$, the approximate and exact solutions
can be evaluated using, respectively, Eq.\ (\ref{approx}) and Eq.\ (\ref{exact}).
{\mgap
The approximate solution in Eq. (\ref{approx}) has an analytic form, 
which is independent of $\alpha$, and is given by \cite{Oli07}
\begin{align}\label{apsol}
\rho(\lambda) = \frac{e^{-\frac12\lambda \sigma_3}\,\tau\,e^{-\frac12\lambda
\sigma_3}}{\hbox{Tr}[\tau\,e^{-\lambda\sigma_3}]}\,,
\end{align}
where $\lambda$ is determined by solving the equation
$\hbox{Tr}[\rho(\lambda)\,\sigma_3]=\langle\sigma_3\rangle$. The
analytic form of the Bloch vector ${\boldsymbol r}=(r_1,r_2,r_3)$  
of $\rho(\lambda) = 
\frac12 \left({\mathbb I}+{\mathbf r}\cdot{\boldsymbol \sigma}\right)$ 
is given by:
\begin{equation*}
r_1 = \frac{t_1}{Z}\ ;\
r_2 = \frac{t_2}{Z}\ ;\
r_3 = \langle \sigma_3 \rangle
\end{equation*}
where ${\boldsymbol t}=(t_1,t_2,t_3)$ is the Bloch vector of the initial 
state $\tau$ and $Z = \cosh{\lambda} - t_3 \sinh{\lambda}$
(see Appendix \ref{a:bvec} for the explicit expression in terms
of the parameters $\theta$, $\phi$, and $\epsilon$). The corresponding
value of the Lagrange multiplier is
\begin{equation}
\lambda = \mathrm{arctanh} \frac{t_3-\langle \sigma_3 
\rangle}{1-\langle \sigma_3 \rangle t_3}\,.
\end{equation}
For what concerns the optimal solution, the first Lagrange multiplier 
$\lambda_1$ is evaluated using the trace normalization for the 
state $\rho$, while the second one $\lambda_2$ is evaluated upon
exploiting the constraint of the mean value. The equation for the last constraint
is transcendental, and numerical methods are
needed in order to find $\lambda_2$.}
When the values of the two solutions are
found, for fixed $\tau$ and $A$, it is possible to compare them using the
qubit fidelity:
\begin{equation}
F(\rho_1,\rho_2)=\tr{\rho_1\rho_2}+\sqrt{1-\mu[\rho_1]}\sqrt{1-\mu[\rho_2]}
\end{equation}
where $\mu[\rho_k]$ is the purity of the state $\rho_k$, $\rho_1$ is the
approximate solution and $\rho_2$ the exact one.
\par
In order to assess the reliability of the approximate solution
we have evaluated the fidelity between the approximate and the
exact solution as a function of the five parameters involved in
the estimation problem. Our first result is that the fidelity
does not depend on the angle $\phi$, i.e. the two solutions
(approximate and exact) show the same functional dependence 
on such parameter. Besides, the approximate and the exact mKE
estimate, as well as the fidelity, do not depend on the parameter 
$\alpha$. 
\par
The relevant parameters to assess the approximate solution are thus 
the angle polar $\theta$ and the purity $\mu[\tau]$ of the initial 
state and the result
of the measurement $\langle \sigma_3 \rangle$. The fidelity is also symmetric 
with respect to the transformations $\theta \rightarrow \pi - \theta$ 
and $\langle \sigma_3 \rangle \rightarrow - \langle \sigma_3 \rangle$.
Notice that finding the exact mKE solution requires the use of numerical 
methods, which pose a upper bound to the initial purity $\mu[\tau] 
\lesssim 1 - 10^{-7}$, above which the solution becomes numerically unstable.
\par 
As we will see in the following, the fidelity shows different behaviors, 
depending on the purity $\mu[\tau]$ of the initial state $\tau$.
Before going to a detailed comparison, we notice that if the
initial state $\tau$ commutes with the measured observable $A$,
i.e.\ $[\tau,A]=0$, then the two solutions coincide as it is
apparent by inspecting Eq.\ (\ref{exact}) and Eq.\ (\ref{approx}).
\subsection{Fidelity for highly mixed initial states}
\begin{figure}[h!]
\centering
\includegraphics[width=0.49\columnwidth]{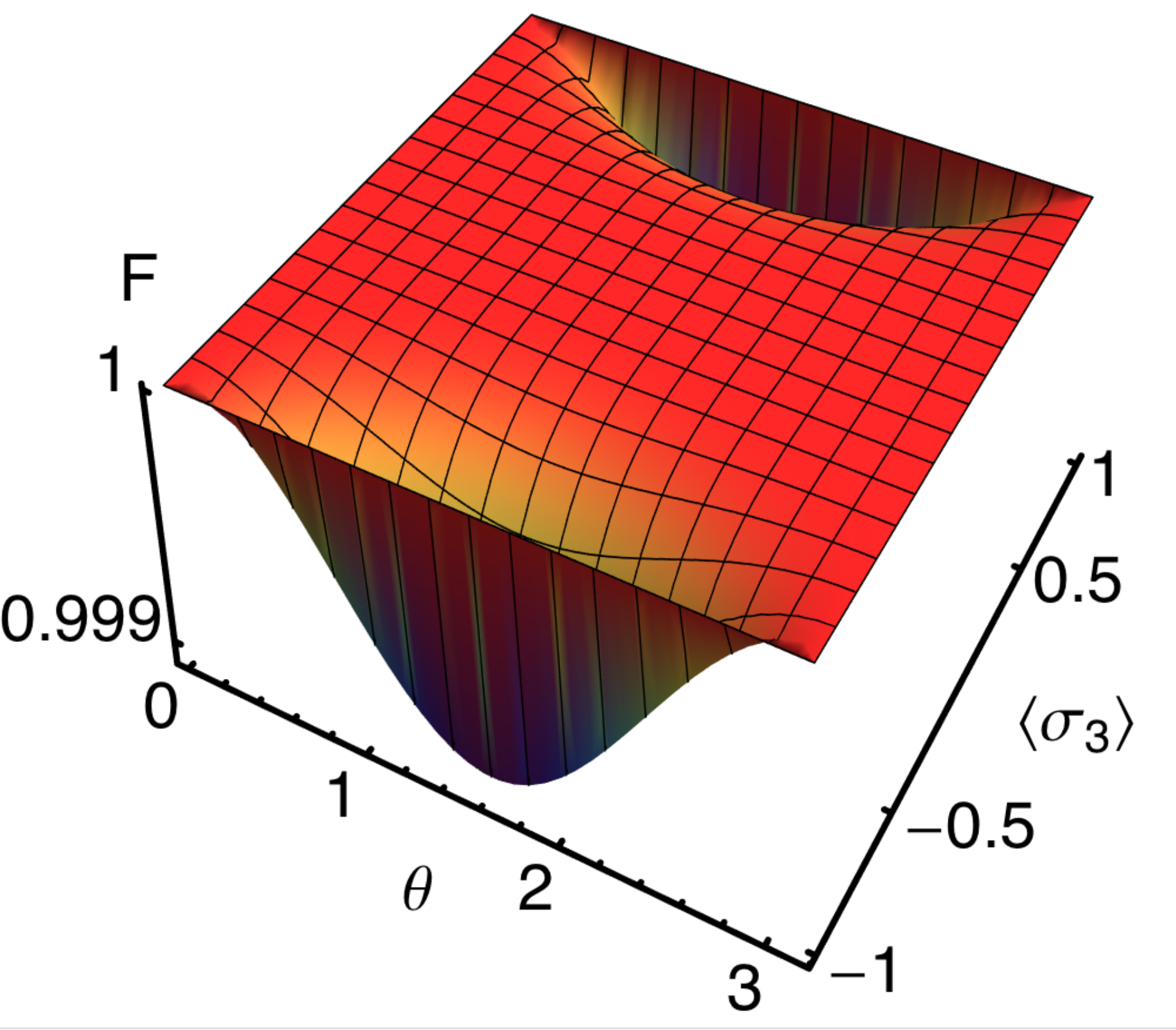}
\includegraphics[width=0.49\columnwidth]{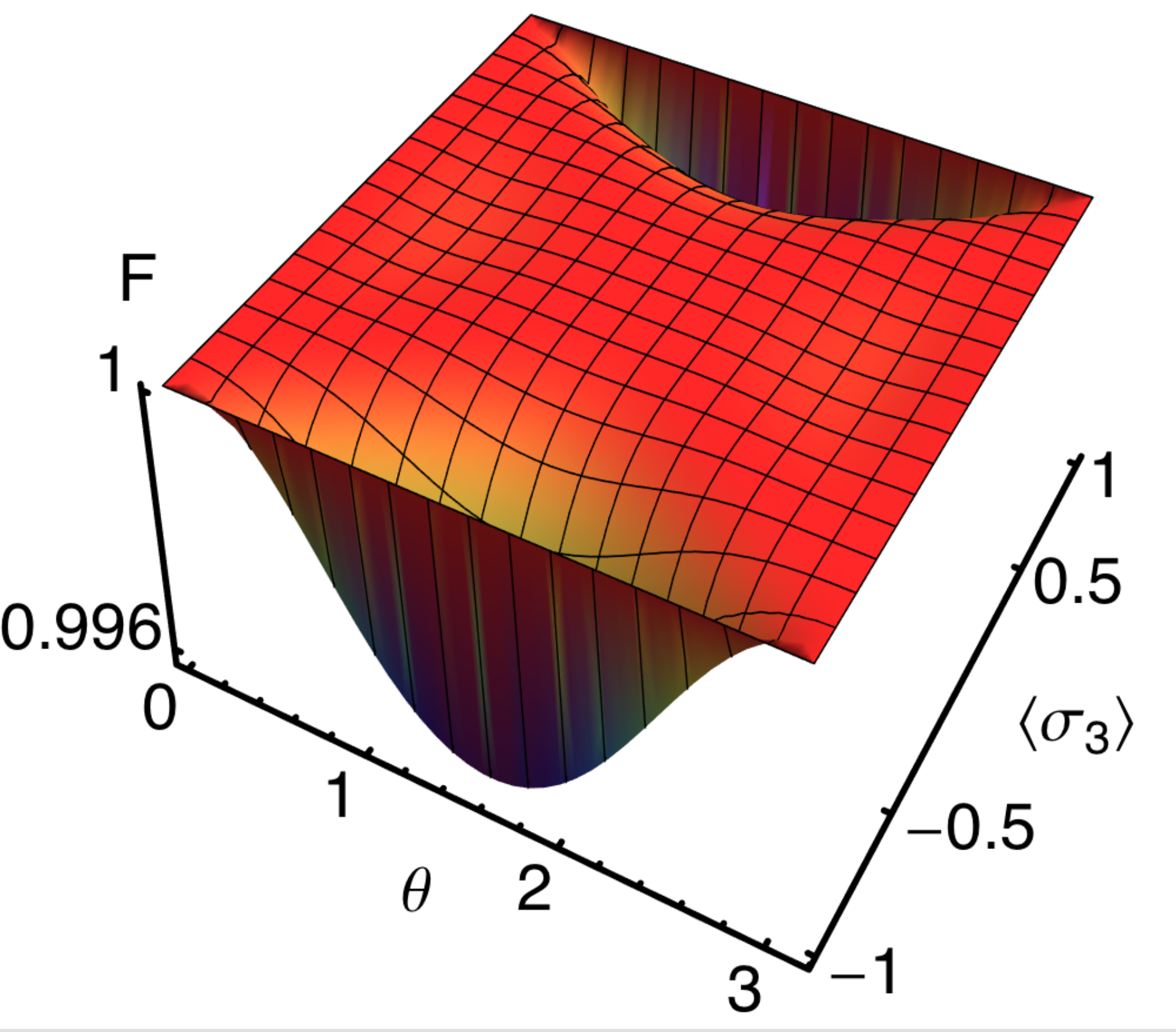}
\includegraphics[width=0.8\columnwidth]{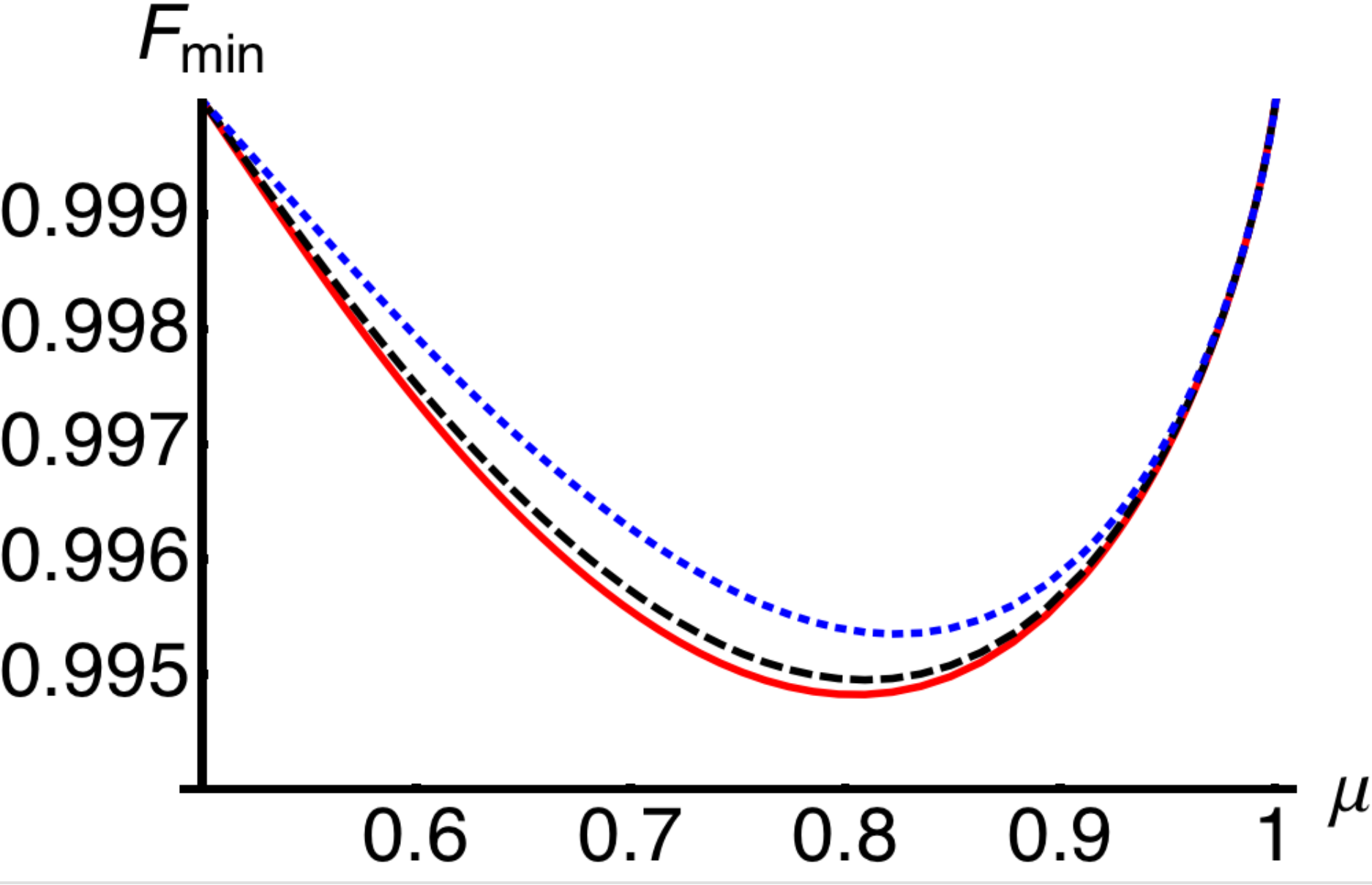}
\caption{(Color online) Fidelity between the exact and the approximate mKE
estimate as a function of the parameter $\theta$ of the initial
state and of the outcome of the measurement $\langle \sigma_3 
\rangle$. The plots are for two fixed values of the initial 
purity: $\mu=0.55$ (top left) and $\mu=0.7$ (top right).
The lower panel shows the minimum of fidelity for
$\theta=\pi/2$ (solid red), $\theta=5\pi/12$ (black dashed), 
$\theta=\pi/3$ (blue dotted) as a function of purity $\mu$.}
\label{f:misti}
\end{figure}

When the purity $\mu$ of the initial state takes values between $1/2$
and, say, $0.9$ (i.e. when the initial state is highly mixed), the 
fidelity presents some distinctive features, highlighted in 
Fig. \ref{f:misti}.
\par
As it is apparent from the plots, the fidelity between the two 
solutions is extremely close to unit for a large range of values 
of $\langle \sigma_3 \rangle$ around $\langle \sigma_3 \rangle=0$, 
whereas for values of $\langle \sigma_3 \rangle$ near $\pm 1$ it 
decreases and shows a minimum. The shape is almost independent on
the value of the initial purity, whereas the minimal value does.
The actual value of the minimum also depends on the angle $\theta$ 
and the global minimum is achieved for $\theta=\pi/2$. It is worth 
noticing that the values of these minima corresponds to 
fidelity always larger than $F=0.995$, i.e. the two estimates are very 
close each other anyway \cite{nota}. 
\par
The values of the global minimum as a 
function of purity, is reported in the lower panel of  Fig.\ \ref{f:misti}.
Notice that when the purity of the state $\tau$ tends to $1/2$, 
i.e. the initial state approaches $\tau=\Id/2$, the two solutions coincides for all 
values of $\theta$ and $\langle \sigma_3 \rangle$, as it may readily seen
from Eq.\ (\ref{exact}) and Eq.\ (\ref{approx}). This corresponds to the
case of a system initially in a maximally mixed state and for which
the measurement is not providing any additional information.
The minimum of the fidelity between the two estimates is observed for
$\mu \simeq 0.8$ and then, for increasing purity, the two solutions 
become again very close each other. 
\begin{figure}[h!]
\centering\label{f3}
\includegraphics[width=0.49\columnwidth]{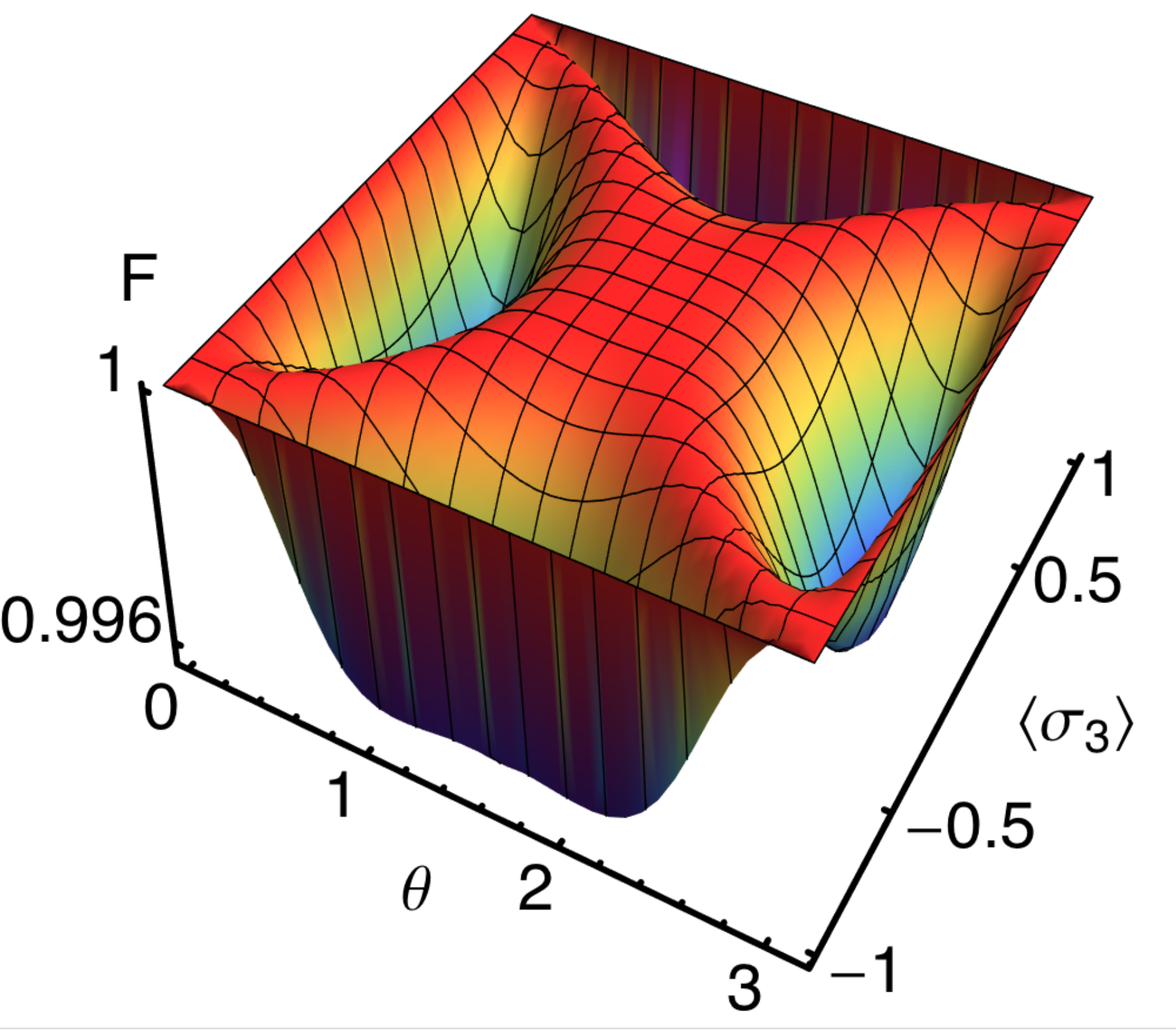}
\includegraphics[width=0.49\columnwidth]{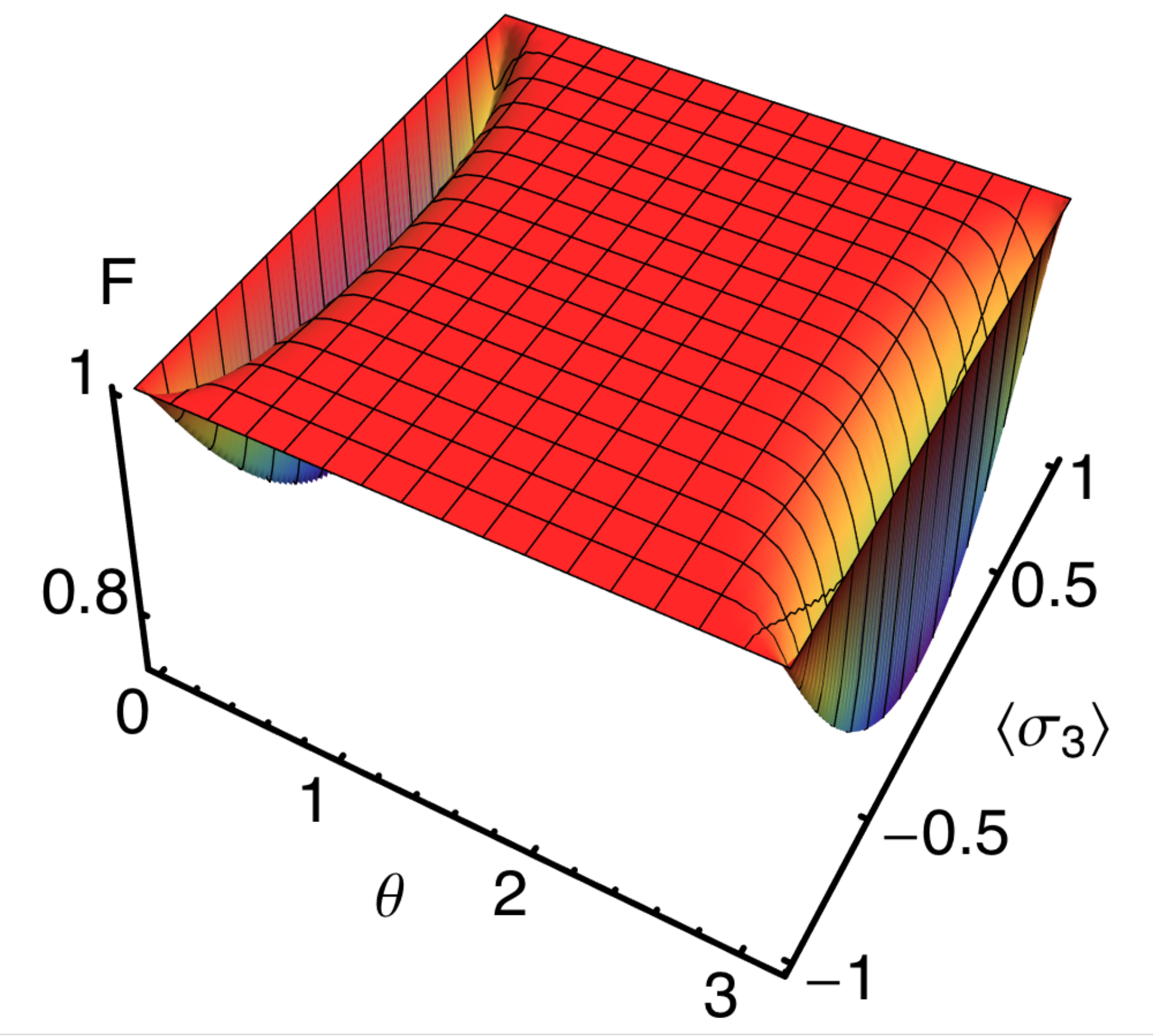}
\includegraphics[width=0.49\columnwidth]{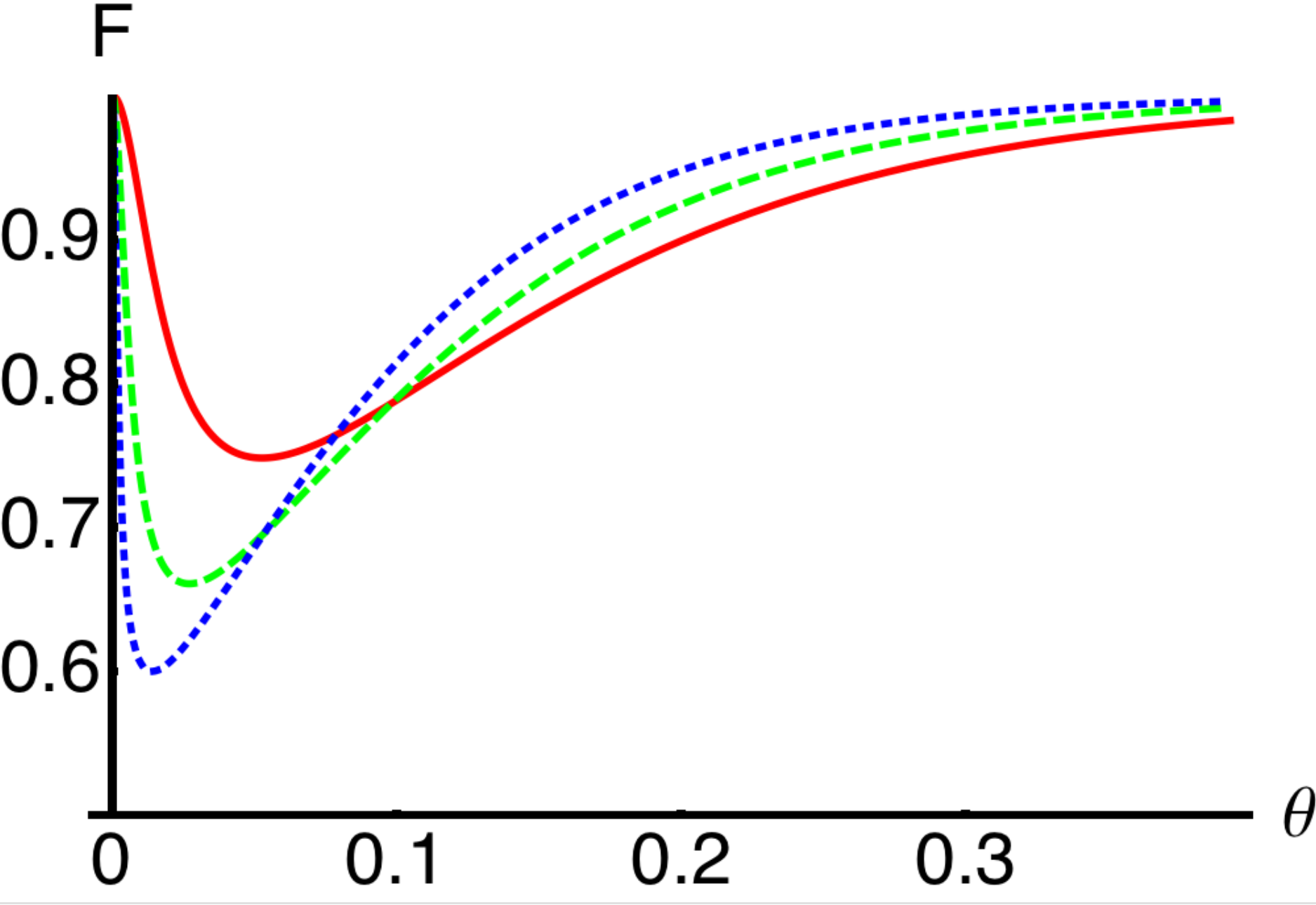}
\includegraphics[width=0.49\columnwidth]{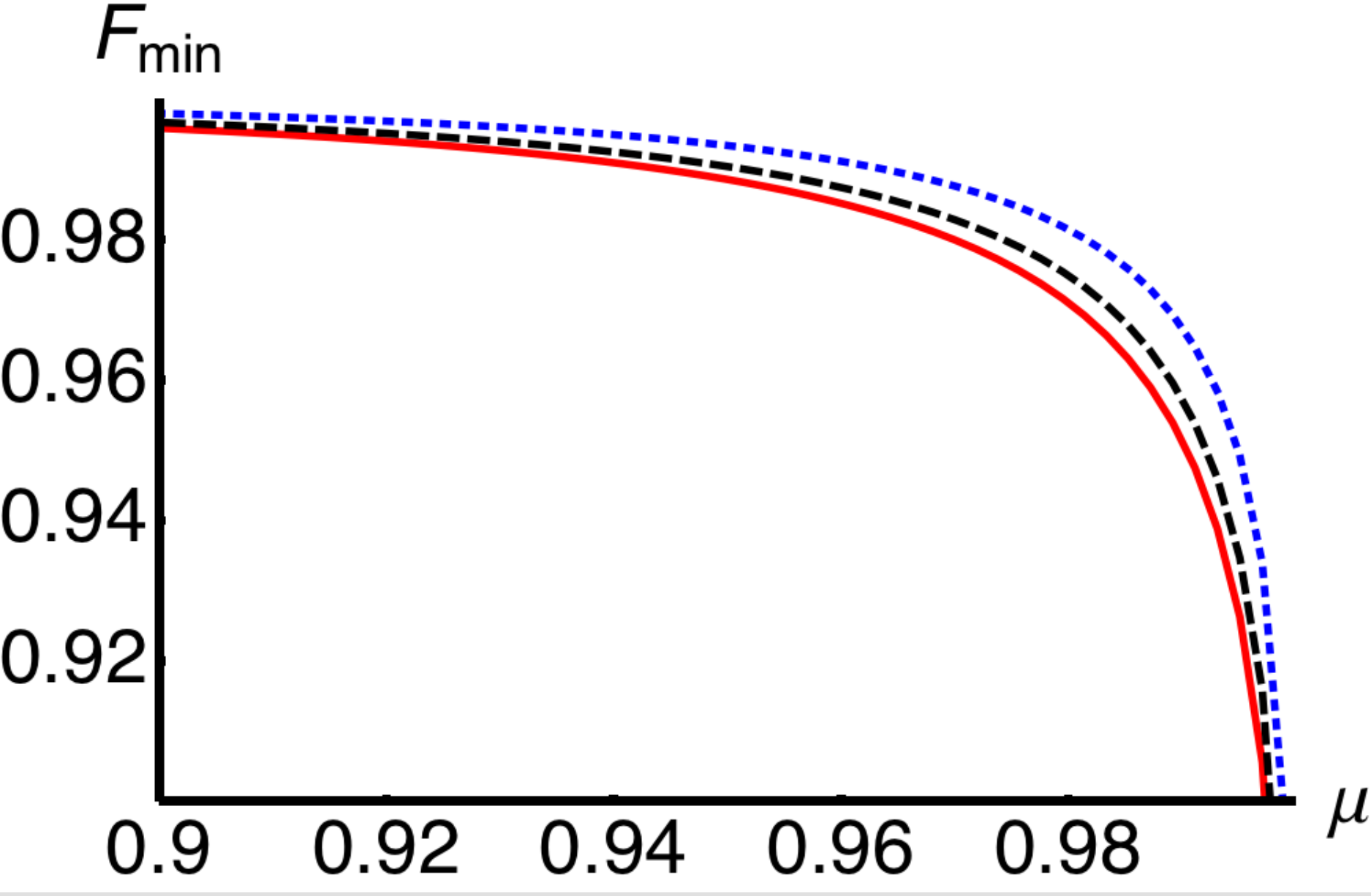}
\caption{(Color online) Fidelity between the exact and the approximate 
mKE estimate for $\mu=0.9$ (left) and $\mu=0.9999$ (right).
The lower left panel shows the behavior of the fidelity 
as a function of $\theta$ for $\langle \sigma_3 \rangle = 0$ and 
for different (close to unit) values of purity: $\mu = 1-10^{-4}$
(solid red line), $\mu = 1-10^{-5}$ (green dashed), $\mu = 1-10^{-6}$ (blue
dotted).
The lower right panel shows the minimum of fidelity for 
for $\langle \sigma_3 \rangle = 0$ (solid red line), 
$\langle \sigma_3 \rangle = \pm 0.3$ (black dashed),
$\langle \sigma_3 \rangle = \pm 0.5$ (blue dotted).
}
\label{f:puri}
\end{figure}
\par
\subsection{Fidelity for nearly pure initial states}
Let us now analyze the situation in which the initial state $\tau$ is closer
to a pure state, with $\mu \in [0.9,1]$. In this regime, the fidelity 
presents a behavior which is quite different from the one illustrated in
the previous Section. In particular, the minima seen for $\theta$ around
$\pi/2$ and $|\langle\sigma_3\rangle|\rightarrow 1$ disappear and are
replaced by minima occurring for $\theta$ close to $0$ or $\pi$ and 
for $\langle\sigma_3\rangle=0$. These phenomena are illustrated 
in the upper panels of Fig. \ref{f:puri}, where we report 
the behavior of fidelity  as a function of $\theta$ and  $\langle\sigma_3\rangle$ 
for two values of the initial purity. The plot for $\mu=0.9$ still shows 
the two kinds of minima, whereas for larger values the transition
from a regime to the other is completed.
\par
As it is apparent from Fig. \ref{f:puri}, the exact and approximate
estimates almost coincide for most values of $\theta$ and
$\langle\sigma_3\rangle$, while their fidelity starts to differ from
unit when $\theta$ is near $0$ or $\pi$.  The discrepancy becomes more
and more appreciable as far as $\langle \sigma_3 \rangle$ approaches
$0$.  Another information extracted from these plots is that the range
of $\theta$ values where the fidelity is appreciably smaller than one
tends to shrink and move towards zero while the purity increases. In
other words, when $\mu \rightarrow 1$, and therefore the initial state
$\tau$ is pure, the minimum of fidelity stays on $\theta = 0$,$\pi$,
while for all the other values of $\theta$ and $\langle \sigma_3
\rangle$ the exact and the approximate mKE solutions coincide. When the
purity of the initial state decreases (still being larger than a
threshold value, say 0.9), we find a neighborhood of $\theta=0$ (and
$\pi$) where the two solutions are different. Furthermore, in this
interval of $\theta$ values, the fidelity decreases towards a minimum,
which approaches to $1/2$ (the two states are completely unrelated) when
$\mu$ goes to unit. 
\par
The lower right panel of Fig. \ref{f:puri} shows the behavior of
the global minimum as a function of purity.
Notice that for a nearly pure initial state, the values taken by the 
fidelity in the minimum may be far from one. This behavior may be
understood as follows: let us consider the point of minimum 
fidelity, i.e. $\theta = 0$ (or $\pi$) and $\langle \sigma_3 \rangle 
= 0$, for $\mu \rightarrow 1$. This corresponds to assume the initial
state $\tau$ to be one of the pure, $\ket{0}$ or $\ket{1}$, 
eigenstates of $\sigma_3$. On the other hand, if the measured mean value 
of $\sigma_3$ is zero this suggest that the state $\rho$ is somehow
mixed. In fact, the exact mKE estimate is the completely mixed state 
$\rho \simeq \Id/2$. On the contrary, the approximate solution is 
the pure state $\ket{\phi}=1/\sqrt{2} (\ket{0} + \ket{1})$. In other
words, the very form of the approximate solution tends to keep the purity of
the initial state unchanged, as it is apparent from  Eq. (\ref{approx})
when we consider $\tau=\ketbra{\psi}{\psi}$.
\begin{figure}[h!]
\centering
\includegraphics[width=0.49\columnwidth]{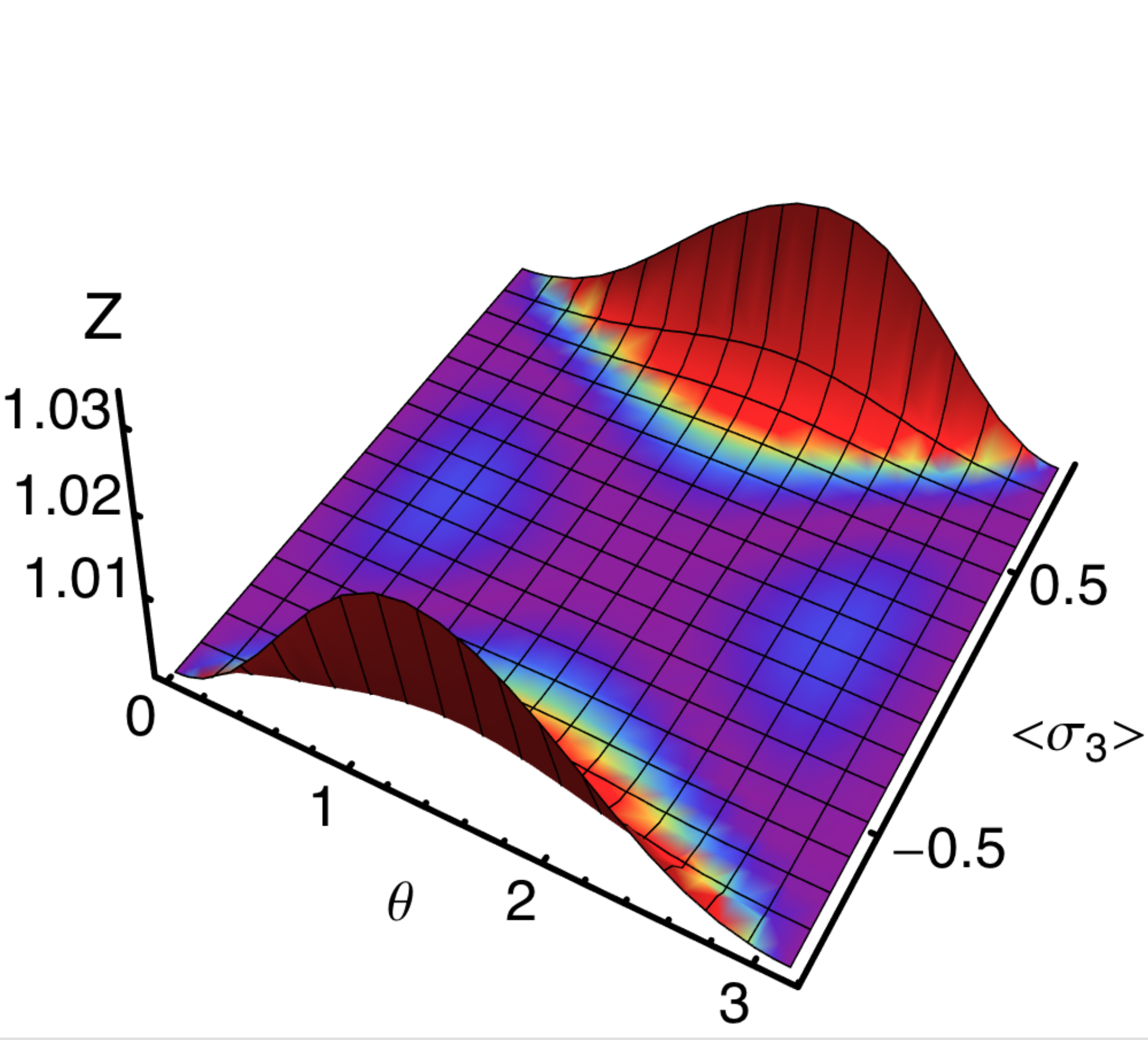}
\includegraphics[width=0.49\columnwidth]{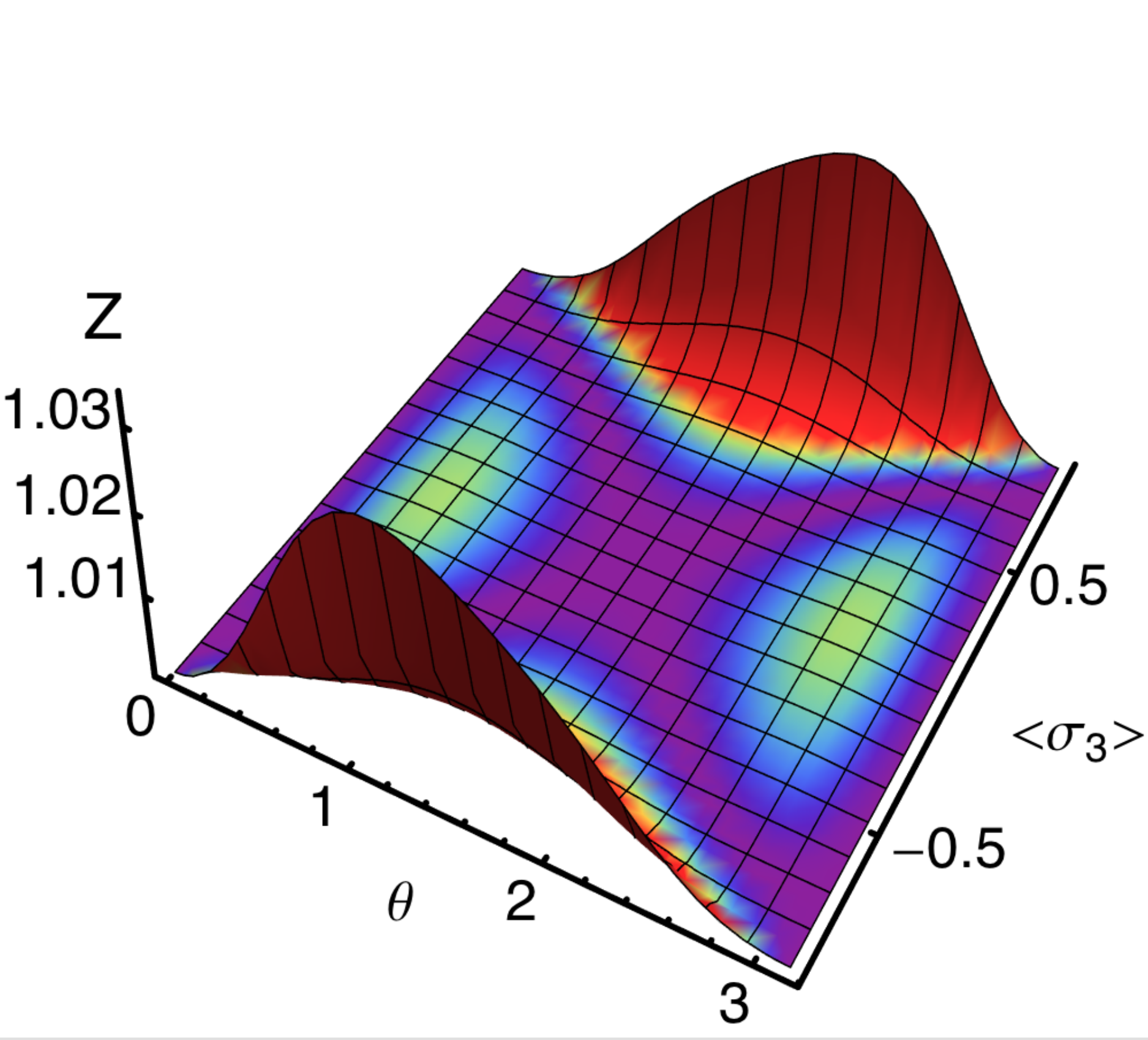}
\includegraphics[width=0.49\columnwidth]{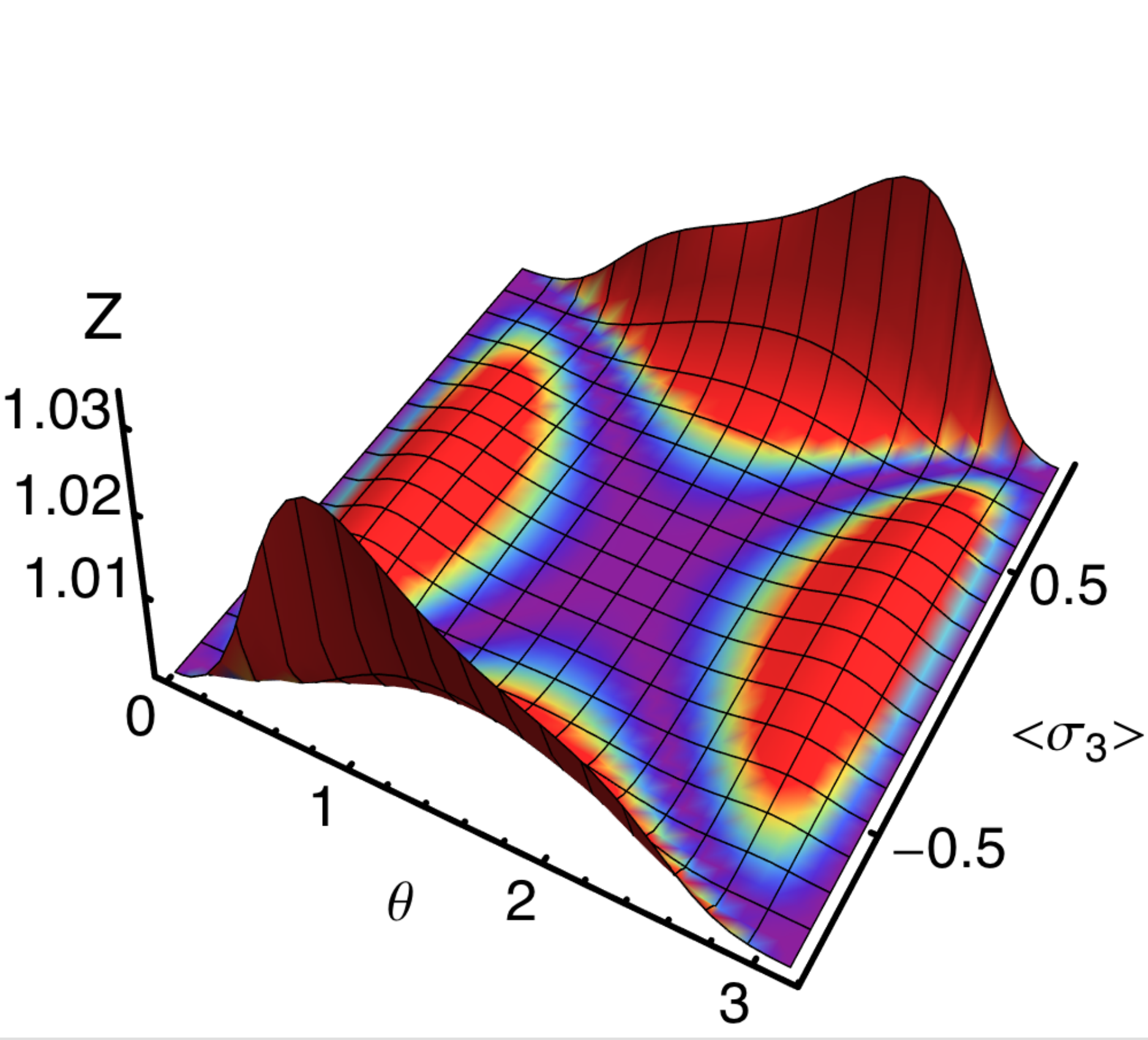}
\includegraphics[width=0.49\columnwidth]{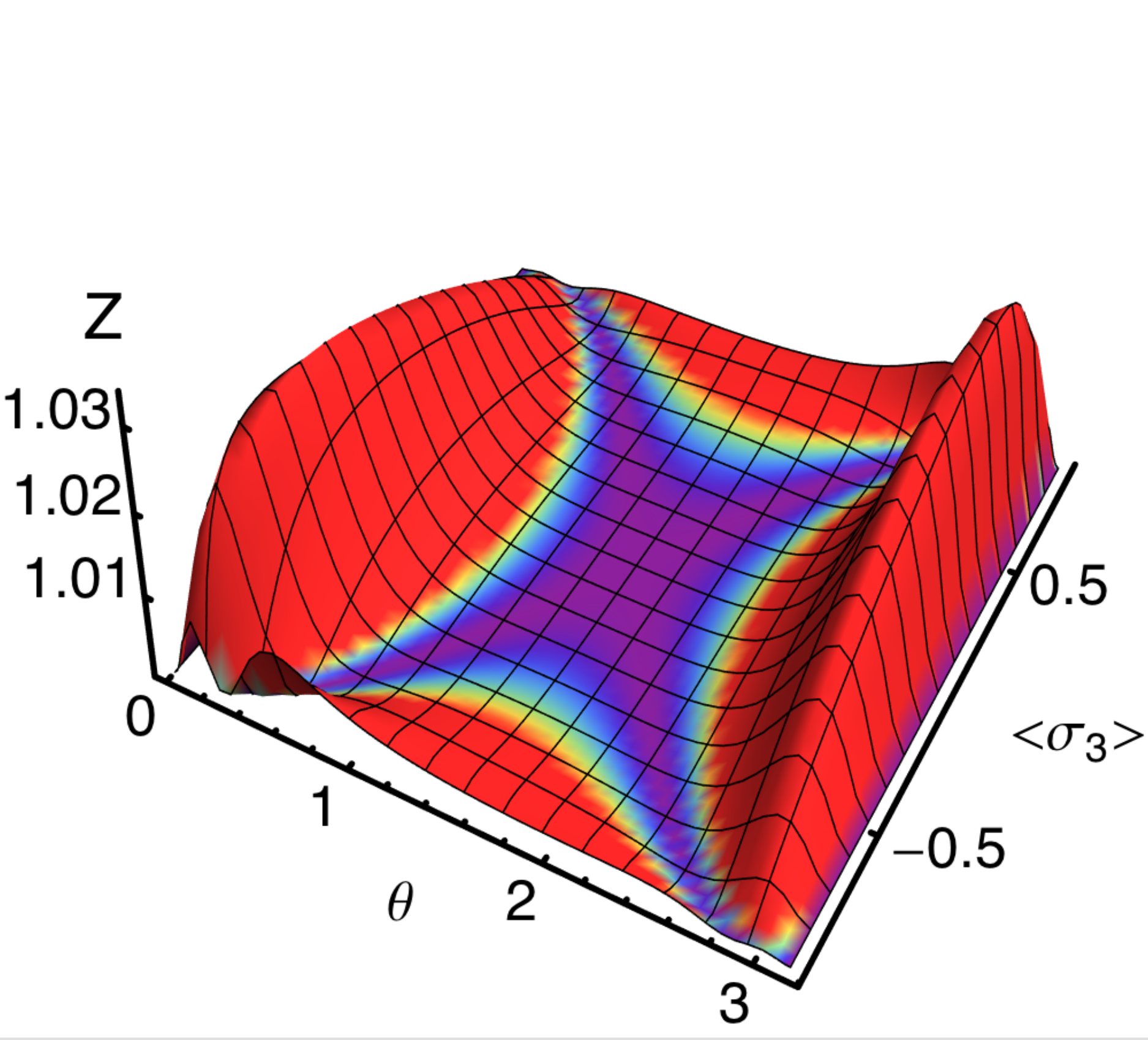}
\caption{(Color online) The ratio, $Z$, between the two fidelities $F(\rho_{apx},\tau)$
and $F(\rho_{exa},\tau)$, as a function of $\theta$ and $\langle \sigma_3 \rangle$,
for different values of initial purity $\mu$.
In the first graphic we have $Z$ for $\mu=0.7$. Then, from top to bottom,
from left to right, we find graphics for $\mu=0.8$, $\mu=0.9$ and $\mu=0.99$.
It is evident that $Z$ is always greater (or equal) than one,
which means that $\rho_{apx}$ is, in general, closer to $\tau$ than $\rho_{exa}$.
}
\label{fid:rapp}
\end{figure}
\par
Actually, the approximate method appears to force the solution to
be closer to the initial state than the exact solution of Eq.
(\ref{exact}) does, in agreement with the assumptions used for
its derivation. More precisely, the approximate solution is closer, in 
terms of fidelity, to the initial state than the exact one, for any 
values of $\mu$, $\theta$ 
and $\langle \sigma_3 \rangle$. This phenomenon is illustrated in Fig. \ref{fid:rapp}, 
where we report the ratio $Z = F(\rho_{apx},\tau)/F(\rho_{exa},\tau)$
between the fidelities of the two solution to the initial state. As 
it is apparent from the plots, we have $Z \geq 1$ for the whole range 
of parameters and, in turn, this confirms the above considerations.
\par
It is worth to notice that both the relative entropy and the fidelity can
be used to measure the similarity between two quantum states. Since the mKE
principle minimizes the Kullback entropy, the exact solution should be closer 
to the initial state than the approximate one in terms of relative
entropy, whereas ther are no constraints on the fidelity. Overall, our
results shows that, in this case, fidelity and relative entropy provides 
two opposite assessments \cite{nota}.
\subsection{The purity of the two solutions}
As we have seen in the previous Sections, when the initial state shows high
purity the approximate solutions tends to preserve such purity
irrespective of the results of the measurements, whereas this 
is not the case for the exact solution. Since this phenomenon
represents the underlying reason of the behavior of fidelity
reported in the previous Section, here we provide a more detailed 
study of the purity of the exact and approximate solutions
as functions of the purity $\mu$ of the initial state $\tau$.
\begin{figure}[h!]
\centering
\includegraphics[width=0.49\columnwidth]{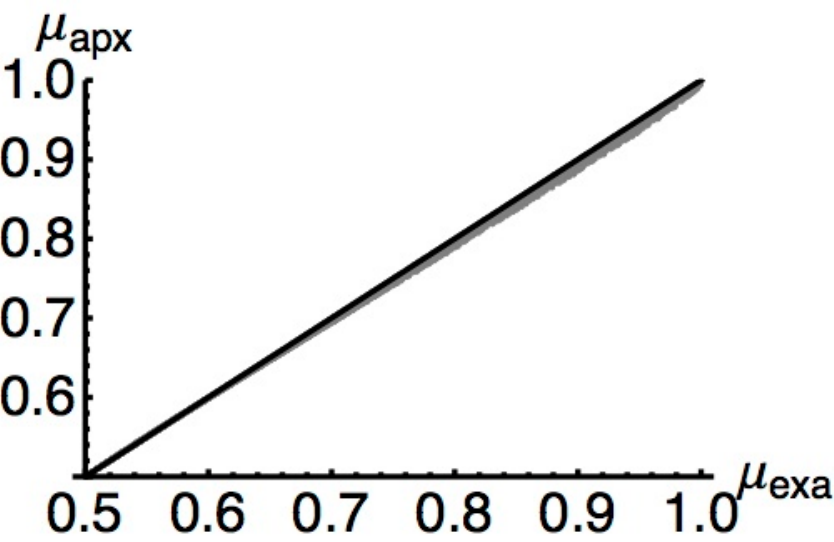}
\includegraphics[width=0.49\columnwidth]{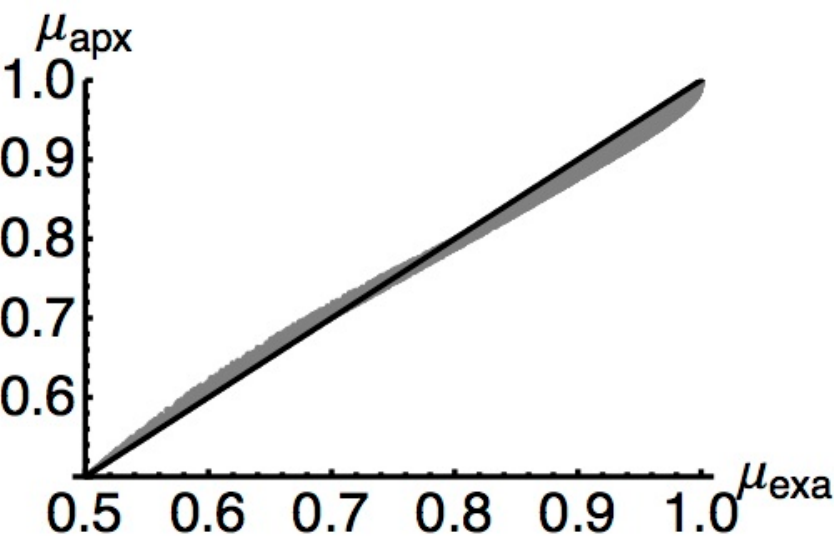}
\includegraphics[width=0.49\columnwidth]{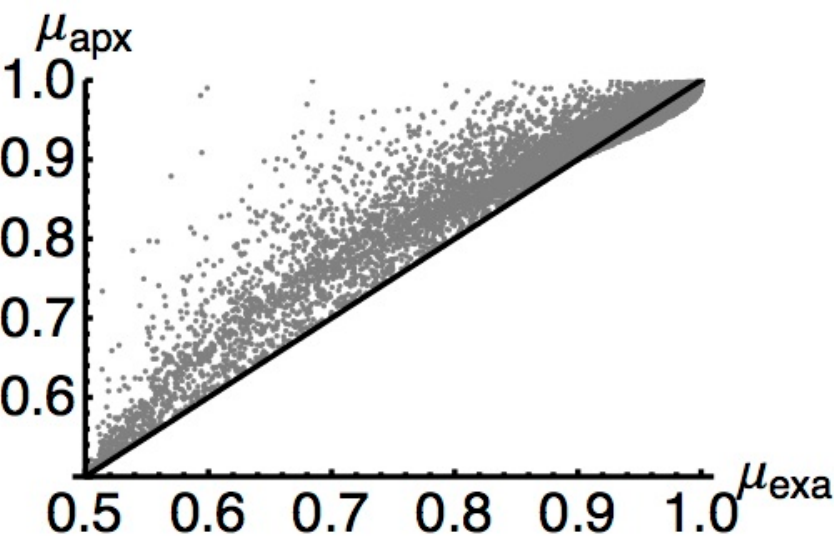}
\includegraphics[width=0.49\columnwidth]{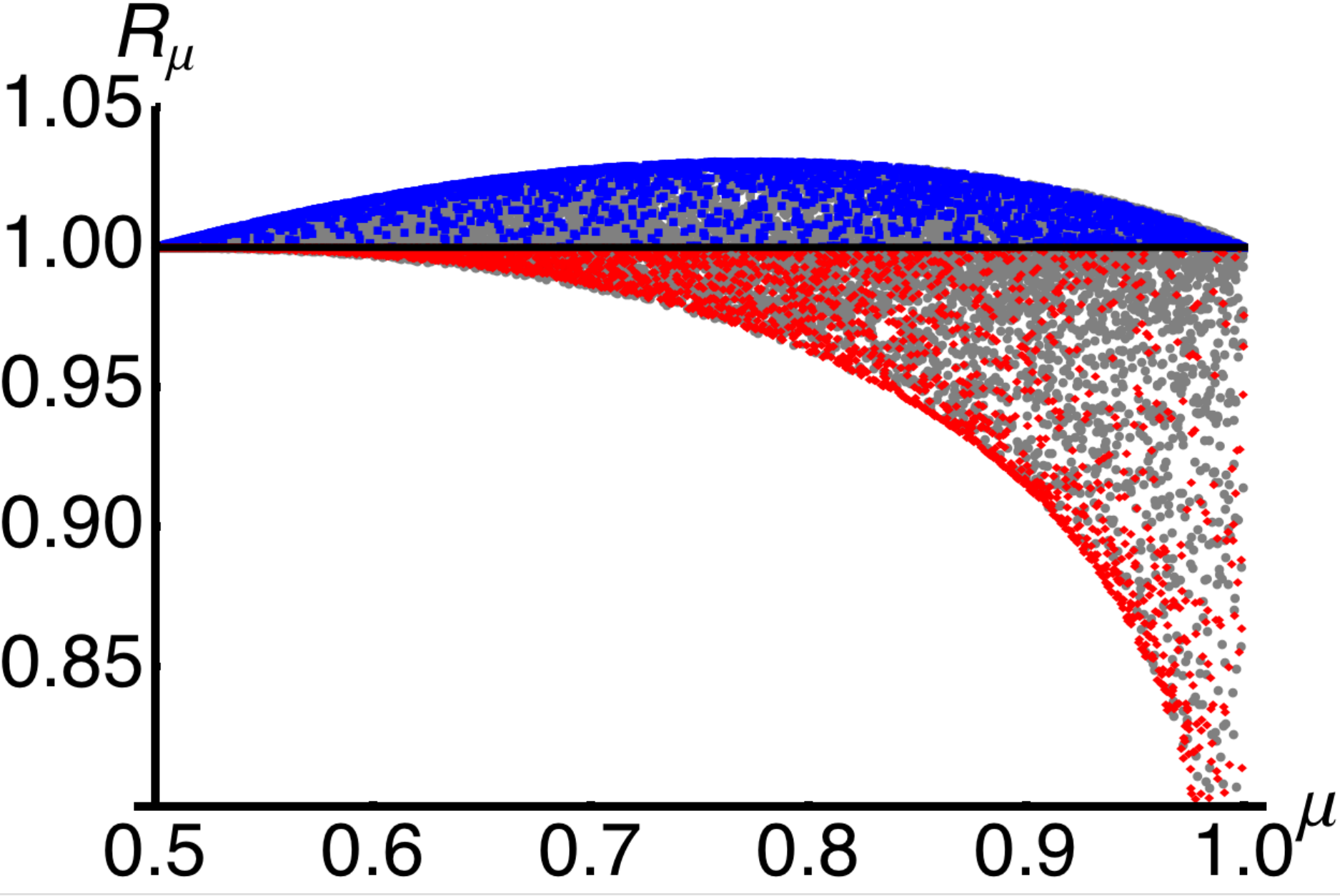}
\caption{(Color online) Purities of the approximate and exact solutions 
of mKE estimation. The first three plots show the purity of the
approximate solutions as a function of the exact one for the initial 
purity in the ranges $\mu=0.5-0.6$ (top left), 
$\mu=0.7-0.8$ (top right) and $\mu=0.9-1.0$ (down left) respectively. 
In all the plots grey points correspond to purities of mKE 
solutions obtained by random selecting the initial states with 
$\theta \in [0,\pi]$ and $\mu$ in the given range, and simulating 
random measurements with $\langle \sigma_3 \rangle \in [-1,1]$. 
The right lower panels shows the ratio $R_{\mu}$ as function of 
initial purity $\mu$. Here the grey points are obtained by randomly 
selecting the initial states in the full range of $\theta$ 
and $\mu$ and $\langle \sigma_3 \rangle \in [-1,1]$, the blue 
(square) points are obtained by sampling $\theta$ in the range $\theta\in
[\pi/2-\pi/10, \pi/2+\pi/10$ and $|\langle \sigma_3 \rangle| \in
[0.9,1]$, and the red (rhombus) points correspond to random sampling 
$\theta\in [-0.01,0.01] $ or around $\pi$ and 
$\langle \sigma_3 \rangle \in [-0.1,0.1]$.}
\label{kb:purity}
\end{figure}
\par
The first three panels of Fig. \ref{kb:purity} report the purity of the
approximate solution $\mu_{apx}$ as a function of the purity of the
exact one $\mu_{exa}$ for randomly generated values of $\theta\in
[0,\pi]$ and $\langle \sigma_3 \rangle \in [-1,1]$. In each plot the
purity $\mu$ of the initial state is randomly sampled in a fixed range:
$\mu\in[0.5,0.6]$ in the upper left plot, $\mu\in[0.7,0.8]$ in the upper
right plot, $\mu\in[0.9,1]$ in the lower left plot. The upper left plot
shows that for highly mixed initial states, the purities of the two
solutions are close each other. For intermediate values of the initial
purity we see a mixed behavior, whereas for the nearly pure initial
states of the lower left plots the approximate solution tends to
preserve their purities, such that $\mu_{apx}$ is larger than
$\mu_{exa}$ for most of the values of $\theta$ and  $\langle \sigma_3 \rangle$.
\par
The plot in the lower right panel of Fig. \ref{kb:purity}
shows the ratio $R_{\mu} = \mu_{exa}/\mu_{apx}$ as a function of the
initial purity $\mu$ for a randomly chosen values of  $\theta$ and
$\langle \sigma_3 \rangle$.  In order to fully appreciate the content
of this plot, let us first consider the case $R_{\mu} > 1$, i.e.
$\mu_{exa}>\mu_{apx}$. This ratio achieves its maximum in the 
region $\mu \in [0.75,0.85]$ and the same behavior may be recognized in the second 
plot of Fig.\ \ref{kb:purity}, where, for high values $\mu_{exa}$ and $\mu_{apx}$, 
we see that $\mu_{exa} > \mu_{apx}$. Notice that, for initial purity $\mu \simeq 0.7$, the
approximate and exact solutions differ only for values of $\theta$ around $\pi/2$ and
$\langle \sigma_3 \rangle$ close to $\pm 1$ (see Fig.\ref{f:misti}).
It thus appears that in this area the exact solution has a purity larger 
than the approximate one. Indeed, this is indeed confirmed by sampling 
$\theta$ around $\pi/2$ and $\langle \sigma_3 \rangle$ near $\pm 1$ 
(blue points in the last plot of Fig.\ \ref{kb:purity}.
For $R_{\mu} < 1$, we have that increasing the initial purity 
corresponds to a decrease of $R_{\mu}$, which achieve its minimum for 
pure initial states. Again, this is due to the fact that the 
approximate solution tends to keep the purity of the initial 
state unchanged, while the exact one does not.
Besides, this behavior may be recognized also in the third plot 
of Fig. \ref{kb:purity}. For initial high purities, the purities of 
the approximate and the exact solutions are different only for values 
of $\theta$ close to $0$ or $\pi$ and $\langle \sigma_3 \rangle$. 
This may be also seen by randomly sampling points in that area, which
corresponds to the red points of the last plot of Fig. 
\ref{kb:purity}.
\section{Comparison between the exact and the approximate mKE estimation of weak Hamiltonians}
\label{s:ham}
As mentioned above, the mKE principle is an useful tool
to estimate the state of a system which has a bias toward
a given state and when some information coming from measurements 
on the final state  are known.
Therefore, this principle may be naturally applied to the
estimation of a weak Hamiltonian, which drives the evolution 
of a system in the neighborhood of the initial state. 
\par
Suppose that a qubit system is described by the initial state $\tau$,
and it evolves according to the Hamiltonian $H$. The state after this
evolution is given by
\begin{equation}\label{stato_evoluto}
\rho = e^{-i H} \tau e^{i H} \,.
\end{equation}
The Hamiltonian can be represented by the 
vector ${\mathbf h}=(h_1, h_2, h_3)$ in the Pauli basis
\begin{equation*}
H = \sum_{j=1}^3 h_j \sigma_j\,.
\end{equation*}
We assume to know the initial state and want to estimate
the Hamiltonian using the information coming from the measurement 
of a single observable $A$ after the evolution. Upon using the 
mKE principle to estimate the output state and expanding Eq. 
(\ref{stato_evoluto}) to the first order in the Hamiltonian strenght  
(in agreement with the hypothesis of weak interaction), 
the estimated coefficients of the Hamiltonian are obtained \cite{Oli07}:
\begin{equation}\label{coefficienti}
{\boldsymbol h} =\frac{{\boldsymbol \tau} \times {\boldsymbol r}}{2 |{\boldsymbol \tau}|^2}
\end{equation}
where ${\boldsymbol \tau}$ and 
${\boldsymbol r}$ are, respectively, the Bloch vectors
of the initial state $\tau = 1/2 (\Id + {\boldsymbol \tau} \cdot {\boldsymbol \sigma})$
and of the final one $\rho = 1/2 (\Id + {\boldsymbol r} \cdot {\boldsymbol \sigma})$.
\par
Since the mKE estimate for the output state may be obtained using 
either the exact method or the approximate one we have two
possible estimates for the Hamiltonian operators, which will be denoted 
by  $H_{exa}$ and $H_{apx}$. As a matter of fact, in both cases the
coefficients are obtained from Eq.\ (\ref{coefficienti}) and thus the
difference between the two Hamiltonians is due to the difference between
the exact and approximate mKE estimates for the states. In other words,
comparing the two Hamiltonians provide a method to compare the two
solutions of the mKE principle in terms of their use as a probe, rather
than in terms of their closeness in the Hilbert space.
\par
\begin{figure}[h!]
\centering
\includegraphics[width=0.49\columnwidth]{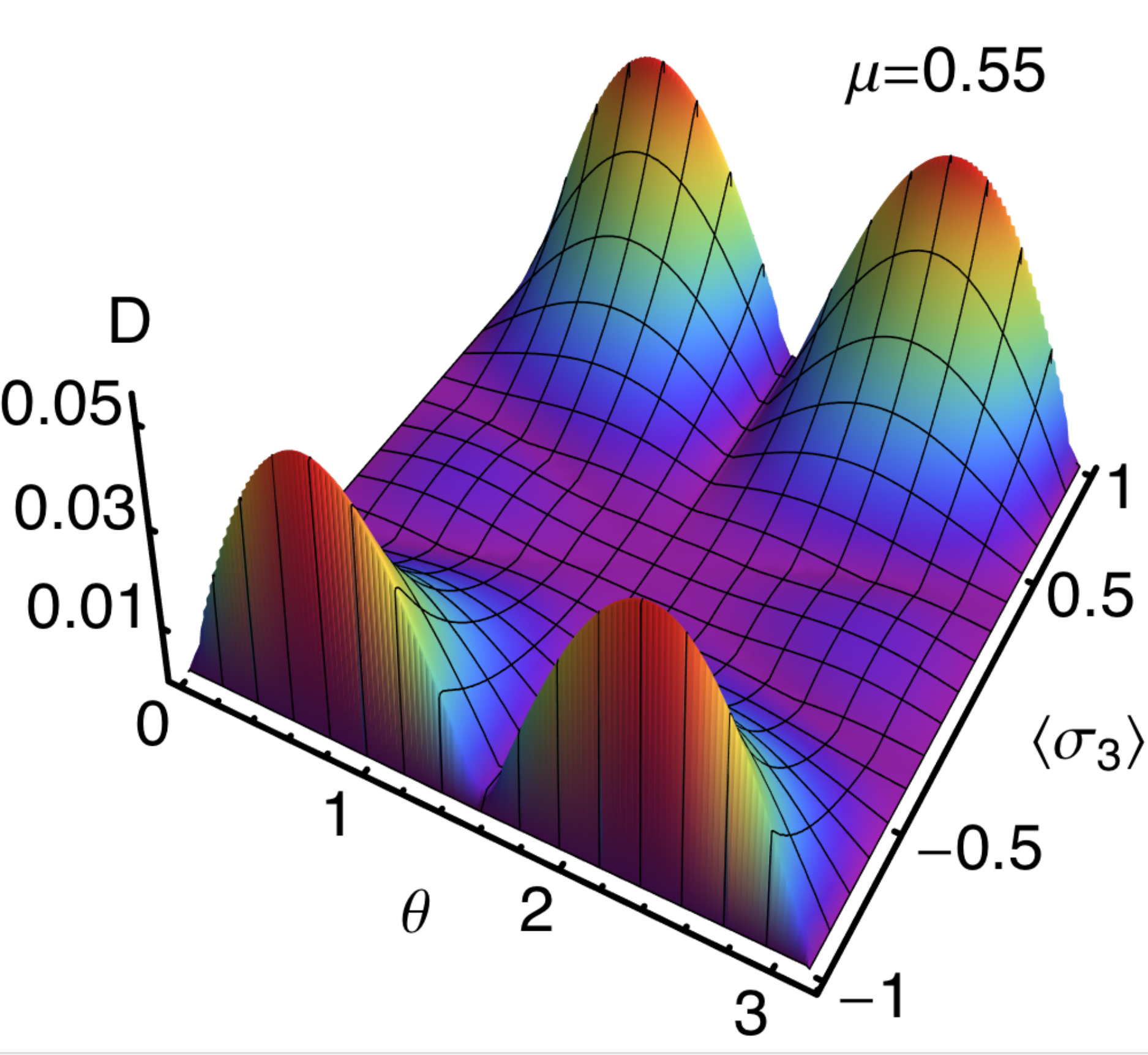}
\includegraphics[width=0.49\columnwidth]{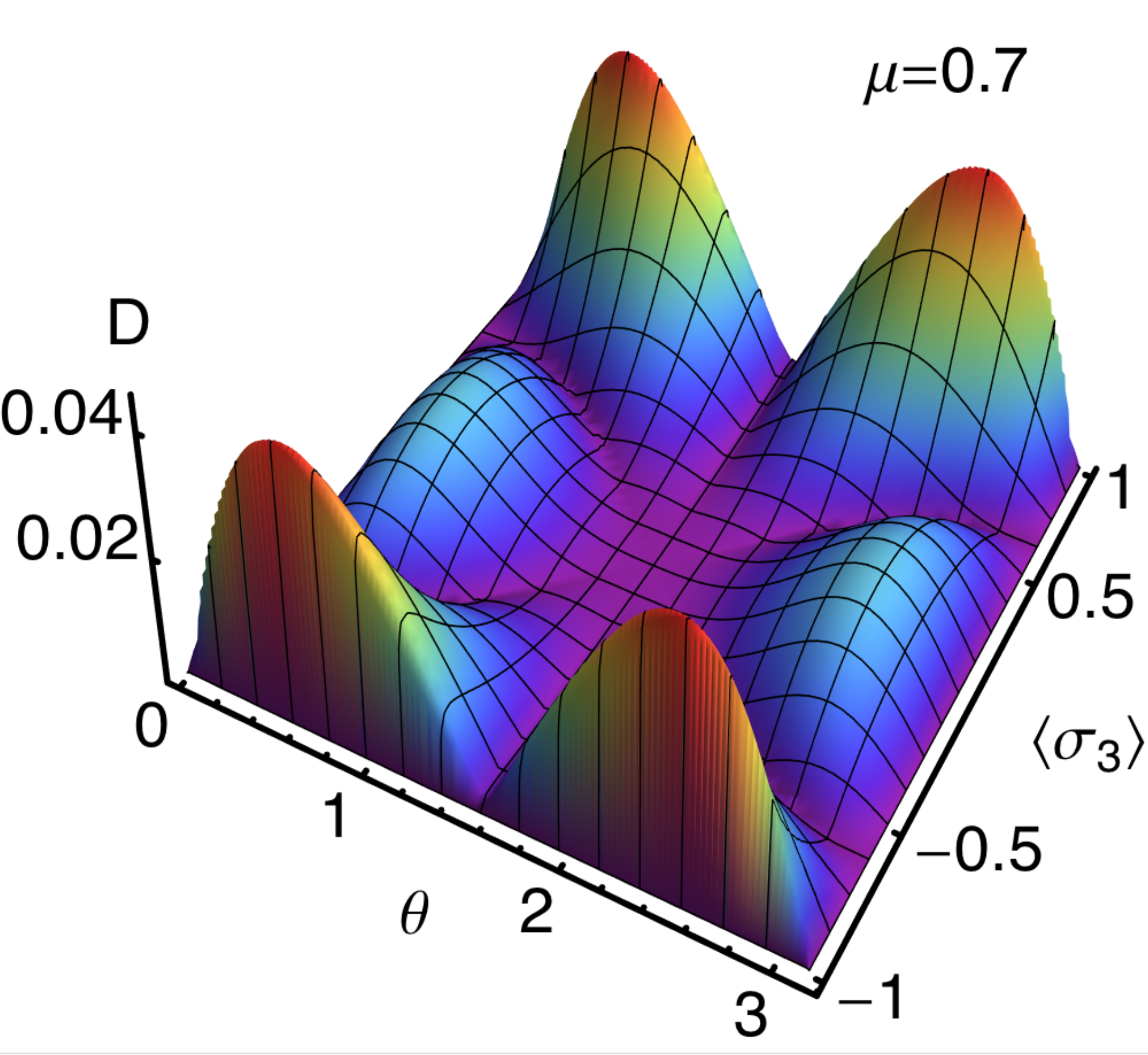}
\caption{(Color online) Comparison between exact and approximate solution
for the mKE estimation of weak hamiltonians.
{\ft On the left side, the trace distance $D$ between the $H_{exa}$
and the $H_{apx}$ is plotted for a purity of the initial state of
 $\mu = 0.55$. On the right, the trace distance $D$ is displayed for $\mu = 0.7$
}}
\label{cnf:misti}
\end{figure}
\par
In order to compare the two Hamiltonians we employ the trace distance,
that is: 
\begin{equation*}
D = \frac{1}{2}\, \tr{|H_{exa}-H_{apx}|}
\end{equation*}
where $|B|$ is the modulus operator of $B$,
i.e. $|B|=\sqrt{B^\dagger B}$. We found that this
quantity does not depend on the phase $\phi$ of the
initial state and shows the same symmetries of 
the fidelity between the two solutions. In the 
following we present a brief analysis of
the behavior of $D$.
\par
When the initial state $\tau$ is highly mixed 
the difference between
$H_{exa}$ and $H_{apx}$ is small, and reaches a maximal value
for $|\langle \sigma_3 \rangle| \rightarrow 1$ and $\theta = \pi/4$ and
$3/4 \pi$ (see Fig.\ \ref{cnf:misti}).
This behavior is similar to the one of the
fidelity for mixed state, but instead of having
a maximal difference for $\theta = \pi/2$, here a minimal difference
is found. This is due to different estimates obtained for the 
components of the Bloch vectors which define the two mKE solutions. In fact,
for $\theta = \pi/2$ and $|\langle \sigma_3 \rangle| \rightarrow 1$,
the first two components (here referred to as $\sigma_1$ and $\sigma_2$)
of the Bloch vectors are small but quite different for the exact
and the approximate mKE estimates. The last component is anyway equal to one for
both the Bloch vectors. The fidelity is able to point this difference 
out, which however is not affecting the estimation of the Hamiltonians, 
since the coefficients of the Hamiltonian focus
only on the larger component of the Bloch vectors. In turn, 
the trace distance between the two estimated Hamiltonians 
do not detect 
a difference between the approximate and the exact method.
The behavior of the distance between the Hamiltonians for 
$\mu \in [1/2,0.7]$ is analogous, though for increasing $\mu$ another
maxima appear, in the same way minima appear for the fidelity.
\par
\begin{figure}[h!]
\centering
\includegraphics[width=0.49\columnwidth]{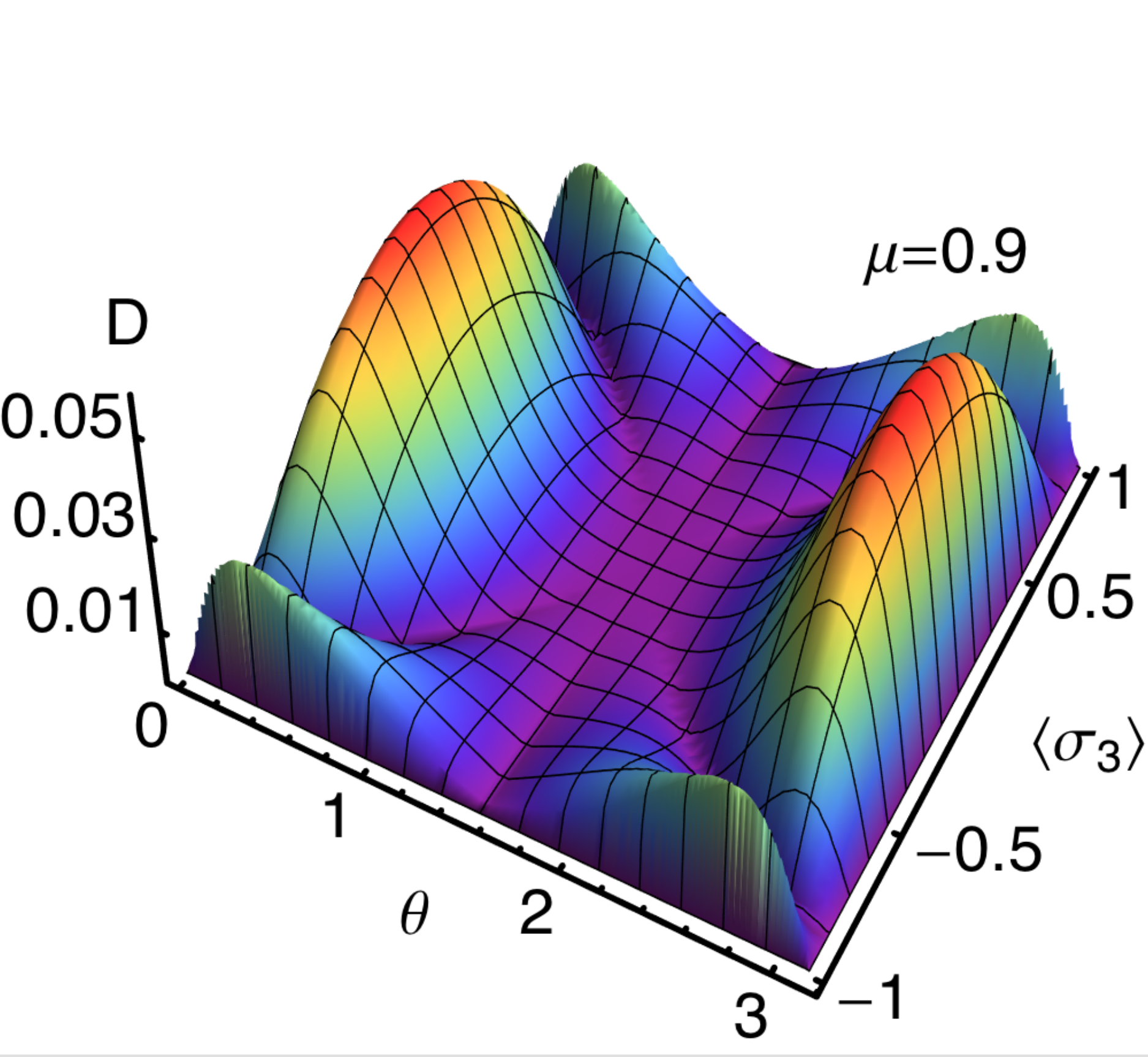}
\includegraphics[width=0.49\columnwidth]{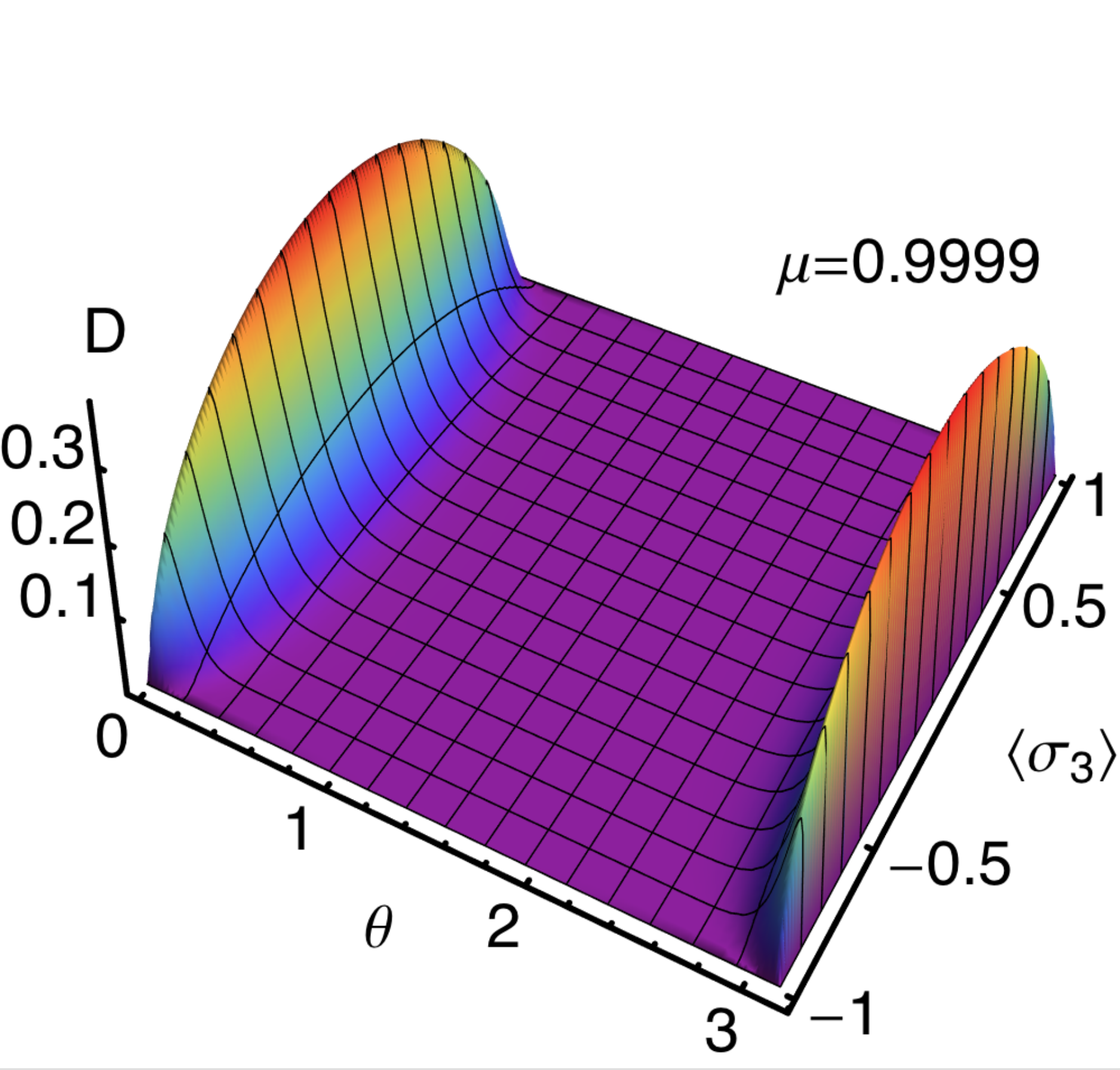}
\caption{(Color online) Comparison between exact and approximate solution
for the mKE estimation of weak hamiltonians.
{\ft On the left side, the trace distance $D$ between the $H_{exa}$
and the $H_{apx}$ is plotted for a purity of the initial state of
$\mu = 0.9$. On the right, the trace distance $D$ is displayed for $\mu = 0.9999$
}}
\label{cnf:puri}
\end{figure}

Let us now address nearly pure initial states: as it is apparent 
from the comparison of Figs.\ref{cnf:misti} and \ref{cnf:puri} 
for increasing $\mu$ we see a transition in the behavior of
the trace distance between Hamiltonians: new maxima appear,
for $\theta \rightarrow 0, \pi$ and $\langle \sigma_3 \rangle = 0$.
and their value increases for $\mu \rightarrow 1$. Overall, 
the behavior of the trace distance $D$ for nearly pure initial 
states is analogous to that of the fidelity in the same regime, 
and all the observations made in that case holds.
\section{Conclusions}
\label{s:out}
In this paper, we have {\ft considered} mKE state estimation for qubits
{\ft when the average of a single observable is available}. In
particular, a detailed comparison between the approximate and exact
solution of the mKE estimation problem has been performed, with the goal
of finding the regimes where the approximate solution may be effectively
employed.  {\mgap In this case, the advantage is that the approximate
solution is given in a closed-form, and it may applied to a larger
class of a priori states, including those described by a density
operator not having not full rank.}
\par
In order to compare the two solutions we have analyzed in details the
behavior of fidelity between the two estimated states as a function of
the parameters of the initial states and of the outcome of the
measurement. Our results show that the most striking difference
concerns the purity of the estimated states, with the approximate
solution that tends to preserve the purity of the initial state, while
the exact one does not, being more sensitive to the information coming
from the measurement outcome. Moreover, we find that in terms of
fidelity the approximate solution is closer to the initial state than the
exact one for the whole range of parameters.  
\par
We have also addressed mKE
principle as a tools to estimate weak Hamiltonians and compared the
performances of the two solutions for this specific task. {\ft Employing the}
trace distance to compare the estimated Hamiltonians, we found results
that confirm those obtained analyzing fidelity.
\par
Overall, {\mgap our analysis} shows that approximate solutions to mKE estimation
problems may be effectively employed to replace the exact ones unless
the initial state is close to an eigenstate of the measured observable.
{\mgap In turns, this provides a rigorous justification for the use of the approximate
solution whenever the above condition does not occur.}
\acknowledgments
This work has been supported by MIUR (project FIRB ``LiCHIS'' - 
RBFR10YQ3H) and the University of Padua (project ``QuantumFuture''). 
\appendix
\begin{widetext}
{\mgap
\section{Bloch vector of the 
approximate solution in Eq. (\ref{apsol})}\label{a:bvec}
Upon solving the equation 
$\hbox{Tr}[\rho(\lambda)\,\sigma_3]=\langle\sigma_3\rangle$, 
we obtain an analytic form for the Lagrange multiplier $\lambda$ and, in
turn, for the Bloch vector of the approximate solution of Eq.
(\ref{apsol}) 
\begin{align}
r_1 &=
\frac{(\epsilon -1) \sin \theta \cos \phi (\langle\sigma_3\rangle (\epsilon -1) \cos
\theta + \epsilon +1) \sqrt{1-\frac{(\langle\sigma_3\rangle (\epsilon +1)+(\epsilon -1)
\cos \theta )^2}{(\langle\sigma_3\rangle (\epsilon -1) \cos \theta +\epsilon
+1)^2}}}{(\epsilon -1)^2 \cos ^2\theta -(\epsilon +1)^2}
\notag \\ 
r_2 &=
-\frac{(\epsilon -1) \sin \theta \sin \phi (\langle\sigma_3\rangle (\epsilon -1) \cos
\theta +\epsilon +1) \sqrt{1-\frac{(\langle\sigma_3\rangle (\epsilon +1)+(\epsilon -1)
\cos \theta )^2}{(\langle\sigma_3\rangle (\epsilon -1) \cos \theta +\epsilon
+1)^2}}}{(\epsilon -1)^2 \cos ^2\theta -(\epsilon +1)^2}
\notag \\ 
r_3 &= \langle\sigma_3\rangle
\,.
\end{align}
}
\end{widetext}

\end{document}